\documentclass[aps,prl,twocolumn,groupedaddress,preprintnumbers,nofootinbib]{revtex4}
\usepackage[dvips]{graphicx}
\usepackage{color}
\usepackage{amsmath,amssymb,slashed}
\usepackage{ulem}
\renewcommand{\[}{\begin{equation}}
\renewcommand{\]}{\end{equation}}

\newif\ifdraft
\drafttrue
\newif\ifpreprint
\preprinttrue
\def\spa#1.#2{\left\langle#1\,#2\right\rangle}
\def\spb#1.#2{\left[#1\,#2\right]}

\def\dd{\hat d}
\def\del{\hat \delta}
\def\qb{\bar q}

\renewcommand{\Re}{\operatorname{Re}}
\renewcommand{\Im}{\operatorname{Im}}
\newcommand{\rootKerr}{$\sqrt{\textrm{Kerr}}$ }

\newcommand{\mF}{\mathcal{F}}

\newcommand{\eq}{\begin{equation}}
\newcommand{\eqe}{\end{equation}}
\newcommand{\eqa}{\begin{eqnarray}}
\newcommand{\eqae}{\end{eqnarray}}

\newbox\charbox
\newbox\slabox
\def\s#1{{      % Feynman slash
        \setbox\charbox=\hbox{$#1$}
        \setbox\slabox=\hbox{$/$}
        \dimen\charbox=\ht\slabox
        \advance\dimen\charbox by -\dp\slabox
        \advance\dimen\charbox by -\ht\charbox
        \advance\dimen\charbox by \dp\charbox
        \divide\dimen\charbox by 2
        \raise-\dimen\charbox\hbox to \wd\charbox{\hss/\hss}
        \llap{$#1$}
}}

\begin{document}

\preprint{NCTS-TH/1905}

\title{Kerr Black Holes as Elementary Particles}

\author{
Nima Arkani-Hamed,$^1$, Yu-tin Huang,$^{2,3}$, Donal O'Connell,$^{4}$}
\affiliation{$^1$ School of Natural Sciences, Institute for Advanced Study, Princeton, NJ 08540, USA}
\affiliation{$^2$ Department of Physics and Astronomy, National Taiwan University, Taipei 10617, Taiwan}
\affiliation{$^3$ Physics Division, National Center for Theoretical Sciences, National Tsing-Hua University,
No.101, Section 2, Kuang-Fu Road, Hsinchu, Taiwan}
\affiliation{$^4$ Higgs Centre for Theoretical Physics, School of Physics and Astronomy, The University of Edinburgh, Edinburgh EH9 3JZ, Scotland, UK}

\begin{abstract}
Long ago, Newman and Janis showed that a complex deformation $z\rightarrow z+i a$ of the Schwarzschild solution produces the Kerr solution. The underlying explanation for this relationship has remained obscure. The complex deformation has an electromagnetic counterpart: by shifting the Coloumb potential, we obtain the EM field of a certain rotating charge distribution which we term $\sqrt{\rm Kerr}$.  In this note, we identify the origin of this shift as arising from the exponentiation of spin operators for the recently defined ``minimally coupled'' three-particle amplitudes of spinning particles coupled to gravity, in the large-spin limit. We demonstrate this by studying the impulse imparted to a test particle in the background of the heavy spinning particle.  We first consider the electromagnetic case, where the impulse due to $\sqrt{\rm Kerr}$ is reproduced by a charged spinning particle; the shift of the Coloumb potential is matched to the exponentiated spin-factor appearing in the amplitude. The known impulse due to the Kerr black hole is then trivially derived from the gravitationally coupled spinning particle via the double copy.
\end{abstract}

\maketitle

%%%%%%%%%%%%%%%%%%%%%%%%%%%%%%%%%%%%%%%%%%%
\section{Introduction}
%%%%%%%%%%%%%%%%%%%%%%%%%%%%%%%%%%%%%%%%%%%
The no hair theorem states that black holes are characterized by only their mass, charge and angular momentum, implying that externally the black hole behaves as a point particle. For a long time this point of view has been utilized to derive the spin-independent part of the two-body classical potential for inspiralling black holes~\cite{Donoghue:1993eb,%
Donoghue:1994dn,Donoghue:1996mt,%
Donoghue:2001qc,BjerrumBohr:2002ks,BjerrumBohr:2002kt,Holstein:2004dn}, from the scattering amplitudes of gravitationally coupled scalars. (See \cite{Cheung:2018wkq, Bern:2019nnu, Foffa:2019yfl, Cristofoli:2019neg} for some recent results,  and \cite{Maybee:2019jus} for a more comprehensive review.)

Of course {\it any} massive object with spin, viewed from sufficiently long distances, can be effectively treated as a point particle. From the perspective of on-shell scattering amplitudes, the most important first issue is to determine the three-particle amplitude, coupling the massive particles to gravitons, and if it is charged, to photons. A convenient on-shell formalism for describing scattering amplitudes for general mass and spin in four dimensions has recently been given in ~\cite{Arkani-Hamed:2017jhn}. In particular the formalism provides a convenient basis for the cubic couplings of massive spin-$S$ particles with a graviton or photon. While for all massless particles of given helicities, three-particle amplitudes are fixed (up to overall strength) by Poincar\'e symmetry, for massive particles of spin $S$ coupled to gravitons or photons, there are $(2S+1)$ different allowed structures, reflecting all the allowed multipole moments of the particle. Returning to the Kerr black hole, this three-particle amplitude coupling to a graviton should be completely prescribed, and is clearly expected to be ``special'' in some way, so the natural question is: what three-particle amplitude is dictated by the no-hair theorem?

From a completely different motivation, ~\cite{Arkani-Hamed:2017jhn} defined a special three-particle amplitude for massive particles of spin $S$ coupled to gravitons and photons, naturally associated ``on-shell'' with a notion of ``minimal coupling'', given by
\eq\label{Minimal}
\vcenter{\hbox{\includegraphics[scale=0.35]{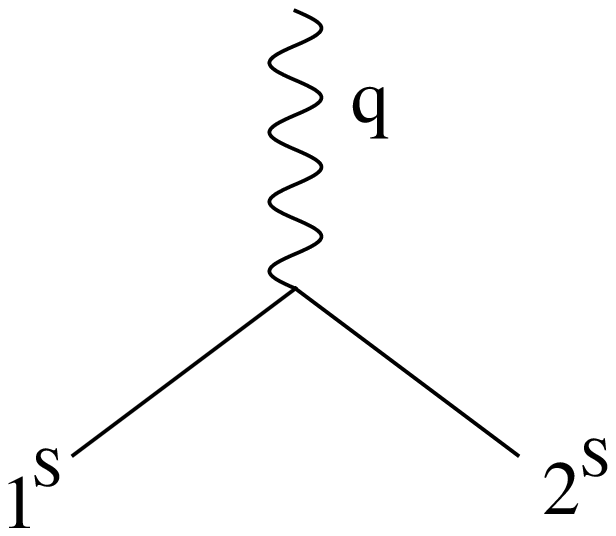}}}\quad= g(xm)^h \frac{\langle \mathbf{1}\mathbf{2}\rangle^{2S}}{m^{2S}}
\eqe
where $h=(1,2)$ and $g=(\frac{\kappa}{2}, \sqrt{2}e)$, for positive photons and gravitons respectively. This coupling was singled out by matching to the (standard, leading) coupling for massless spin $S$ particles in the high energy limit. Indeed for low spins, this coupling reproduces all the classical electric and magnetic moments.

We therefore have a three-particle amplitude picked out as being special purely from the on-shell perspective, making the massive particle look as ``elementary'' as possible to the graviton/photon probe by correctly matching the high-energy limit. Meanwhile, we also know that the Kerr black hole must make a very special choice for the three-particle amplitude as well. Remarkably, the minimally coupled amplitudes are indeed precisely the ones enjoyed by Kerr black holes. Following the work of Guevara ~\cite{Guevara:2017csg}, it was shown in \cite{Guevara:2018wpp} and \cite{BH} that the potential for Kerr black holes was indeed reproduced relativistically to all orders in the multipole expansion from minimal coupling.

These results establish the equivalence of the minimal coupling in eq.(\ref{Minimal}) and Kerr black holes in the context of classical observables, 
but {\it why} did this happen? In this note we would like to give a more fundamental understanding of why minimally coupled higher-spin particles at large spin correspond to Kerr black holes. We will do this by relating minimal coupling to some classic features of the Kerr solution.

Not long after Kerr wrote down the solution for spinning black holes~\cite{Kerr:1963ud}, Newman and Janis observed that one can ``rederive" the Kerr metric by complexifying the Schwarzschild solution in null polar coordinates and performing a shift~\cite{Newman:1965tw}. The construction was later extended to a derivation of the Kerr-Newman solution from Reissner-Nordstrom~\cite{Adamo:2014baa}. For other solutions derived in a similar fashion, see~\cite{Burinskii:1974zz}. The methods of amplitudes allow us to understand the origin of the complex shift. We will demonstrate that the shift is a consequence of the spin effects generated when one goes from a minimally coupled scalar to a spinning particle. In particular starting with a spin-$S$ particle and taking the classical limit, $S\rightarrow \infty, \hbar \rightarrow 0$ while keeping $\hbar S$ fixed, the minimal coupling exponentiates~\cite{Guevara:2018wpp}. This exponent can be identified as $\frac{q\cdot s}{m}$, where $s^\mu$ is the Pauli-Lubanski pseudovector, $q^\mu$ the massless momentum and $m$ the mass. When applied to the computation of  classical observables, such as the change in momentum a probe experiences in a gravitational or electromagnetic field, this exponentiation precisely induces the relevant shift, after Fourier transforming to position space. In other words, the exponentiation incurred going from minimally-coupled scalars to spinning particles, is the momentum space image of the complex shift that relates the Schwarzschild to the Kerr solution. This sharpens the equivalence between black holes and particles.

This connection also provides an on-shell realization of the double copy relation for classical solutions.  In an earlier work by one of the authors~\cite{Monteiro:2014cda}, it was shown that stationary
Kerr-Schild metrics admit a double copy construction. In particular the double and single copy solutions take the form:
\eq
g_{\mu\nu}=g^{0}_{\mu\nu}+k_\mu k_\nu \phi(r),\quad A^{\mu a}= c^a k^\mu \phi(r)\,,
\eqe
where $\phi(r)$ is the universal part for the gravity/gauge theory solution and $k_\mu$ a null $(r,\theta)$-dependent vector. Passing from the $\phi(r)$ for Schwarzschild to Kerr, one simply takes a complex shift. On the other hand, as discussed previously, the difference between the three-point amplitude for gravitational and electromagnetic minimal coupling is simply the squaring of the $x$ factor, whilst the spin-dependent part is untouched. The later corresponds to the shifted $\phi(r)$, while the squaring can be identified with the squaring of $k_\nu$. To illustrate this, we compute the impulse for $\sqrt{\rm Kerr}$ and match it to that from the minimally coupled charged spinning particle. One then obtains the gravitational counterpart by squaring all $x$-factors, which simply translate to a factor of two in rapidity. Remarkably this simple factor of two converts the electromagnetic impulse to the gravitational version.
%%%%%%%%%%%%%%%%%%%%%%%%%%%%%%%%%%%%%%%%%%%
\section{Complexifying Schwarzschild and the double copy}
%%%%%%%%%%%%%%%%%%%%%%%%%%%%%%%%%%%%%%%%%%%
 An early example of the utility of complexified space-time was the derivation of the Kerr metric from a complex coordinate transformation of the Schwarzschild metric~\cite{Newman:1965tw}. We will make use of the metric in Kerr-Schild form:
 \eq
 g_{\mu\nu}=g^{0}_{\mu\nu}+k_\mu k_\nu \phi \,,
 \eqe
where $g^{0}_{\mu\nu}$ is the flat Minkowski metric, and the vector $k_\mu$ is null with respect to both $g^{\mu\nu}$ and $g^{0,\mu\nu}$. In particular, the Schwarzschild solution takes the form
\eq
{\rm Schwarzschild}:\quad \phi_{\rm Sch}(r)=\frac{r_0}{r},\quad k^\mu \partial_\mu=\frac{\partial}{\partial t} -\frac{\partial}{\partial r}
\eqe
where $r_0 = 2GM$. For the Kerr solution, one instead has
\eq
{\rm Kerr}:\quad \phi_{\rm Kerr}(r)=\frac{r_0r}{r^2+a^2\cos^2\theta},\quad k^\mu\partial_\mu=\frac{\partial}{\partial t} -\frac{\partial}{\partial r} \,.
\eqe
Unlike the Schwarzschild case, for Kerr $(r,\theta)$ are not the usual polar coordinates but are defined by:
\eqa
x&{=}&(r\sin\phi{+}a\cos\phi)\sin\theta, \;y{=}(a\sin\phi{-}r\cos\phi)\sin\theta\nonumber\\
z&{=}&r\cos\theta\,.
\eqae
In particular, in the Kerr case $r$ is the solution to the equation
\[
\frac{x^2+y^2}{r^2 + a^2} + \frac{z^2}{r^2} = 1 \,.
\label{eq:spheroids}
\]

It is remarkable that $\phi_{\rm Kerr}$ can be obtained from $\phi_{\rm Sch}$ by a complex shift, which is as simple as $z \rightarrow z + ia$. To see how this connects the Schwarzschild to the Kerr solution, note that the quantity $r^2 = x^2 + y^2 + z^2$ shifts to $x^2 + y^2 + z^2 - a^2 + 2 i a z = r^2 - a^2 \cos^2 \theta + 2 i a r \cos \theta = (r+ia \cos \theta)^2$, where now $r$ is the solution to equation~\eqref{eq:spheroids}. In short, the replacement $z \rightarrow z + ia$ is equivalent to the replacement $r \rightarrow r + i a \cos \theta$. The action on $\phi(r)$ is
\eqa
\left.\phi_{\rm Sch}(r)\right|_{r\rightarrow r+ia \cos\theta}=\left.\frac{r_0}{2}\left(\frac{1}{r}+\frac{1}{\bar{r}}\right)\right|_{r\rightarrow r+ia \cos\theta}\nonumber\\
=\frac{r_0r }{r^2+a^2\cos^2\theta}=\phi_{\rm Kerr}(r)\,.
\eqae
Indeed it is straightforward to show that the Riemann tensors of the two solution are related via this complex shift~\cite{Newman:1965tw}. We are unaware of any classical understanding of why this remarkably simple procedure should work; however, we will see that it follows directly from the nature of observables computed from on-shell amplitudes.

The Kerr-Schild form of the metric is particularly convenient for revealing double copy relations between classical solutions of the Yang-Mills and Einstein equations. It was previously shown~\cite{Monteiro:2014cda} that for every stationary Kerr-Schild solution to the Einstein equations, i.e. $\partial_0 \phi=\partial_0 k_\mu=0$, one finds a solution to the Yang-Mills equation with
\eq
A^{\mu a}= c^a k^\mu \phi(r)\,.
\eqe
For example consider the Schwarzschild case $\phi(r)=\frac{r_0}{r}$. Using the replacement $r_0\rightarrow g c_aT^a$, one finds the static Coulomb potential after a suitable gauge transformation. On the other hand, beginning with the Coulomb charge but performing a complex coordinate shift, one finds the electromagnetic field of a rotating disc with radius $a$~\cite{Monteiro:2014cda}. This Yang-Mills solution is the ``square root" of the Kerr solution, and therefore we call it $\sqrt{\rm Kerr}$. In fact, \rootKerr was discussed by Newman and Janis~\cite{Newman:1965tw} as a complex deformation of Coulomb, and also more recently by Lynden-Bell~\cite{LyndenBell:2002dr}. It correspond to the EM field of Kerr-Newman  where both $M$ and $S$ are sent to zero while holding $a$ fixed. In the following we will compute the impulse probe particles incur in this background and relate the results to Kerr.

%%%%%%%%%%%%%%%%%%%%%%%%%%%%%%%%%%%%%%%%%%%
\section{From $\sqrt{\rm Kerr}$ to spinning particles}
%%%%%%%%%%%%%%%%%%%%%%%%%%%%%%%%%%%%%%%%%%%
We first study the equivalence between the electromagnetic field of the $\sqrt{\rm Kerr}$ solution with the minimally coupled spinning particle, in the infinite spin limit, by computing the impulse induced on a charged particle. In the process we will identify the Kerr parameter $a$ with $\frac{s}{m}$, where $s$ and $m$ are the absolute value of the spin-vector and mass of the particle, respectively.

%%%%%%%%%%%%%%%%%%%%%%%%%%%%%%%%%%%%%%%%%%%
\subsection{Impulse from $\sqrt{\rm Kerr}$ }
%%%%%%%%%%%%%%%%%%%%%%%%%%%%%%%%%%%%%%%%%%%
Performing the complex shift $z\rightarrow z+ia$ on the Coloumb electric field $E_c$, we obtain a complex quantity, $E_c\rightarrow \mathcal{E}$. The interpretation is simple: $\Re \mathcal{E}$ is the electric field of $\sqrt{\textrm{Kerr}}$, while $\Im \mathcal{E}$ is the magnetic field. Covariantly, the complex shift induces a complex field strength $F^{\mu\nu}$. The Lorentz force on a particle with mass $m$, momentum $p^\mu$ and proper velocity $u^\mu$ moving under the influence of the \rootKerr fields is
\begin{equation}
\frac{dp^\mu}{d\tau}
= e \Re \mathcal{F}^{\mu\nu} u_\nu \,,
\label{eq:niceLorentz}
\end{equation}
where $\mathcal{F}^{\mu\nu}=F^{\mu\nu}+i \epsilon^{\mu\nu\rho\sigma}F_{\rho\sigma}/2$.

In electrodynamics, the field strength is gauge invariant and observable. However this fact already fails for Yang-Mills theories, and therefore it is desirable to understand these classical solutions from a different, more gauge invariant point of view. To that end we consider the impulse, that is the total change of momentum, from past infinity to future infinity, of a light particle (particle 1) moving in the (very heavy) \rootKerr background. The impulse on the particle
is computed via:
\eq
\Delta p_1^\mu=e_1 {\rm Re} \int^{\infty}_{-\infty}d\tau\;\; \mathcal{F}^{\mu\nu}(x_1) u_{1\nu}\,.
\eqe
This impulse can be computed perturbatively by iterating the Lorentz force. At leading order, the trajectory of particle 1 is simply a straight line, which is also the all-order trajectory of the source---which we take to be particle 2. Thus for both particles we have $x_i(\tau)=b_i+u_i\tau$, where $u_i$ is the proper velocity; while $b_1$ is real and $b_2$ is complex, reflecting the $\sqrt{\rm Kerr}$ nature of particle 2. Without loss of generality we set $b_1=0$, and $b_2={-}b {-} i a$ so that $b_1 {-} b_2 = b {+} i a$. It is convenient to work in Fourier space, with the field strength due to the source written as
\begin{align}
\mF_2^{\mu\nu}(x_1) = \int \! \dd^4 \qb \, \mF_2^{\mu\nu}(\qb) \, e^{-i \qb \cdot x_1}.
\end{align}
Our notation is that $\qb$ is a wavenumber (momentum transfer $q$ with a $\hbar$ scaled out), $\dd \qb \equiv d \qb / (2\pi)$ and $\del(x) = (2\pi) \delta(x)$. One then has:
\begin{align}
\Delta &p_1^\mu = e_1 \Re\int \! \dd^4 \qb \,  \mF_2^{\mu\nu}(\qb) u_{1\nu} \,  \del(\qb \cdot u_1) \nonumber\\
=& e_1 \Re\int \! \dd^4 \qb  \left( F_2^{\mu\nu}(\qb) {+} \frac{i}2 \epsilon^{\mu\nu\alpha\beta} F_{2 \alpha\beta}(\qb) \right)
% \nonumber\\ & \qquad \qquad \qquad \times
 u_{1\nu}\del(\qb \cdot u_1).
\end{align}
We need an expression for the field strength $F$ in Fourier space. Using the Maxwell equation
\begin{align}
\partial^2 A_2^\mu(x) = e_2 \int d\tau u_2^\mu \delta^4 (x-x_2(\tau)) \,,
\end{align}
it's easy to see that, to all orders for static \rootKerr\!,
\begin{align}
F_2^{\mu\nu}(\qb) = i e_2 \, e^{-i \qb \cdot {(b+ia)}} \del(\qb \cdot u_2) \frac{1}{\qb^2} (\qb^\mu u_2^\nu - \qb^\nu u_2^\mu).
\end{align}
With this information, we obtain our final expression for the impulse in momentum space:
\begin{align}\label{eq:FTimpulse}
\Delta p_1^\mu =& e_1 e_2 \Re \int \! \dd^4 \qb \,   \del(\qb \cdot u_1) \del(\qb \cdot u_2)\frac{e^{-i \qb \cdot {(b+ia)}}}{\qb^2} \left( i \qb^\mu u_1 \cdot u_2 \right.\nonumber\\
&\left.{+}\epsilon^{\mu\nu\alpha\beta} \qb_\nu u_{1\alpha}  u_{2\beta} \right).
\end{align}
Note that the presence of the Levi-Civita tensor is a reflection of the complexification of the field strength.
We will now reproduce the above result from the scattering amplitude involving minimally coupled spinning particles.

%%%%%%%%%%%%%%%%%%%%%%%%%%%%%%%%%%%%%%%%%%%%%%
\subsection{Impulse from $x$}
\begin{figure}
\begin{center}
\includegraphics[scale=0.5]{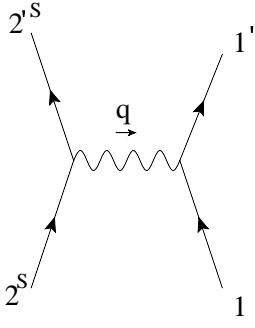}
\caption{The exchange of a photon between a spin-$S$ and a scalar particle. }
\label{fig1}
\end{center}
\end{figure}

%%%%%%%%%%%%%%%%%%%%%%%%%%%%%%%%%%%%%%%%%%%%%%
The impulse for scalar particles was computed from amplitudes in~\cite{Kosower:2018adc} via:
\begin{align}\label{eq:impulseGeneral}
\Delta p_1^\mu =\frac{1}{ 4m_1m_2} \int \dd^4 &\qb\;\del(\qb\cdot u_1)\del(\qb\cdot u_2)e^{{-}i\qb\cdot b}  \times \nonumber \\
&i\qb^\mu \, \hbar^3 M_4\left(1,2\rightarrow 1',2'\right)|_{\qb^2\rightarrow 0} \,
\end{align}
where $M_4$ correspond to a four point amplitude exchanging gravitons or a photons with momentum transfer $q$.  As we will see using this prescription we indeed reproduce the correct impulse for the $\sqrt{\rm{Kerr}}$ electromagnetic field in eq.(\ref{eq:FTimpulse}), by the scattering of a scalar particle 1 with the minimally coupled spin-$S$ particle 2, illustrated in fig.\ref{fig1}.

When dressed with the external polarization tensors, the three-point, minimally-coupled amplitude is given as~\cite{Arkani-Hamed:2017jhn}
\eq
%\vcenter{\hbox{\includegraphics[scale=0.6]{3ptamp}}}\quad
h={+}1:\;\; \sqrt{2} ie_2x\frac{\langle \mathbf{2}\mathbf{2}'\rangle^{2S}}{m^{2S{-}1}},\quad h={-}1:\;\; \sqrt{2} ie_2\frac{1}{x}\frac{[\mathbf{2}\mathbf{2}']^{2S}}{m^{2S{-}1}} \,.
\eqe
Since $q$ is small, the spinor $|\mathbf{2}'\rangle$ is only a small boost of the spinor $| \mathbf{2} \rangle$. We may therefore write
\[
| \mathbf{2}' \rangle = |\mathbf{2} \rangle + \frac{1}{8} \omega_{\mu\nu} (\sigma^\mu \bar \sigma^\nu - \sigma^\nu \bar \sigma^\mu) |\mathbf{2} \rangle \,,
\]
where the boost parameters $\omega_{\mu\nu}$ are small. It is easy to compute these boost parameters because
\[
p_2'^\mu = (\delta^\mu_\nu + \omega^\mu{}_\nu) p_2^\nu \implies \omega_{\mu\nu} = - \frac{1}{m_2^2} (p_{2\mu} q_\nu - p_{2\nu} q_\mu) \,,
\]
taking account of the on-shell relation $2p_2\cdot q = q^2 \simeq 0$. We therefore learn that
\[
| \mathbf{2}' \rangle = |\mathbf{2} \rangle + \frac1{2m_2} \slashed{q} |\mathbf{2} ].
\]
Thus, we have,
\[
\frac{1}{m_2} \langle \mathbf{2} \mathbf{2}' \rangle = \mathbb{I} + \frac1{2m_2^2} \hbar \langle \mathbf{2} |  \slashed{\qb} |\mathbf{2} ] =\mathbb{I} + \frac{1}{2S m_2} \qb \cdot s \, ,
\]
where $s^\mu$ is the Pauli-Lubanski pseudovector associated with a spin $S$ particle:
\[
s^\mu =  \frac{1}{m_2} S \hbar \langle \mathbf{2} | \sigma^\mu | \mathbf{2}] \,.
\]
The operators $\mathbb{I}$ and $s^\mu$ are now operators acting on the little group space of particle $2$. In the end, all little group indices will be contracted with appropriate wave functions. We now take the limit $S \rightarrow \infty$ and $\hbar \rightarrow 0$ with $S\hbar$ fixed. The amplitudes become
\begin{align}
h={\pm}1:&\;\; \lim_{S\to \infty} {i}{e_2}{\sqrt{2}}m\,x^{{\pm}1} \left(\mathbb{I}{\pm} \frac{\qb \cdot s}{2Sm}  \right)^{2S} = {i}{e_2}{\sqrt{2}}m x^{{\pm}1} e^{{\pm}\qb \cdot a}
\label{eq:RootKerrAmplitude}
\end{align}
where the quantity $a = \frac{s}{m}$ parameterises the spin, but has dimensions of length.

Now let's consider the classical limit of the four point amplitude between a charged particle of spin $S$, with $S \rightarrow \infty$, and a scalar particle:
\[\label{Factor}
M_4\left(1,2{\rightarrow} 1',2'\right)|_{\footnotesize q^2{\rightarrow} 0}
=2\frac{e_1e_2 {m_1 m_2}}{2\qb^2}\left(\frac{x_{11'}}{x_{22'}}e^{{-}\qb\cdot {a}}{+}\frac{x_{22'}}{x_{11'}}e^{\qb\cdot {a}}\right)
\]
Note that it is given by two terms with different helicity configurations. We will see that these terms have a crucial role, allowing us to understand the emergence of the real-part operation in the impulse. The $x$ ratios are little group invariant, and can be shown to be given by:
\begin{align}
\frac{x_{11'}}{x_{22'}}=  e^w, \quad\frac{x_{22'}}{x_{11'}}=  e^{-w}\,,
\end{align}
where $w$ is the rapidity. Thus we have,
\eqa
M_4\left(1,2{\rightarrow} 1',2'\right)|_{q^2\rightarrow 0}={2}\frac{e_1e_2 {m_1 m_2}}{\hbar^3 \qb^2}\left(e^w e^{{-}\qb\cdot {a}}{+}e^{{-}w} e^{\qb\cdot {a}}\right)\nonumber\\
\eqae
We now proceed to compute the impulse by inserting this four-point amplitude into the general expression~eq.(\ref{eq:impulseGeneral}):
\begin{align}\label{Rapid}
\Delta p_1^\mu &=i \frac{e_1 e_2}{2}\int \! \dd^4\qb\; \del(\qb\cdot u_1)\del(\qb\cdot u_2)e^{{-}i\qb\cdot b}  \frac{\qb^\mu }{\qb^2}\nonumber\\
&\quad\quad \sum_{\alpha=\pm}e^{\alpha(w{-}\qb\cdot{a})}\,.
\end{align}
To proceed, it's helpful to rewrite the impulse as
\begin{align}
\Delta p_1^\mu &=i \frac{e_1 e_2}{2}\int \! \hat{d}^4\qb\; \hat{\delta}(\qb\cdot u_1)\hat{\delta}(\qb\cdot u_2)e^{{-}i\qb\cdot b} \frac{\qb^\mu }{\qb^2} \nonumber\\
&\left((\cosh w{+}\sinh w) e^{{-}\qb\cdot\Pi {a}}{+}(\cosh w{-}\sinh w) e^{\qb\cdot{a}}\right)
\end{align}
Note that on the support of $ \hat{\delta}(\qb\cdot u_1)\hat{\delta}(\qb\cdot u_2)$, the Gram determinant constraint takes the form 
\begin{align}
\epsilon(u_1, u_2, a, \qb)^2 =-\sinh^2w (a \cdot \qb)^2 + \mathcal{O}(\qb^2).
\end{align}
We may neglect any terms of order $\qb^2$, because in the Fourier integral such terms lead to a delta function in impact parameter space. Furthermore, using a ``Schouten'' identity we have
\begin{align}
\qb_\mu \epsilon(u_1, u_2, a, \qb) = (a \cdot \qb)\epsilon_{\mu \nu \alpha \beta} \qb^\nu u_1^\alpha u_2^\beta   \, .
\end{align}
where $u_1 \cdot \qb$, $u_2 \cdot \qb$, $\qb^2$ are all set to zero. Thus we can identify:
\begin{align}
\sinh w\, \qb_\mu = i\epsilon(u_1, u_2, a, \qb) \frac{1}{a\cdot \qb} \qb_\mu = i\epsilon_{\mu\nu\alpha\beta} \qb^\nu u_1^\alpha u_2^\beta
\end{align}
With this result, the impulse is then:
\begin{align}\label{Impulse1}
\Delta p_1^\mu &=\frac{e_1 e_2}{2}\int \! \hat{d}^4\qb\; \hat{\delta}(\qb\cdot u_1)\hat{\delta}(\qb\cdot u_2)\frac{i}{\qb^2}  \nonumber\\
&\left[(\qb^\mu\cosh w{+}i\epsilon^{\mu\nu\alpha\beta} \qb_\nu u_{1\alpha} u_{2\beta}) e^{-i\qb\cdot( b{-}i {a})}\right.\nonumber\\
&\left.{+}(\qb^\mu\cosh w{-}i\epsilon^{\mu\nu\alpha\beta} \qb_\nu u_{1\alpha} u_{2\beta}) e^{-i\qb\cdot( b{+}i {a})}\right]\nonumber\\
&= e_1 e_2\Re\int \! \hat{d}^4\qb\; \hat{\delta}(\qb\cdot u_1)\hat{\delta}(\qb\cdot u_2)\frac{i}{\qb^2}  \nonumber\\
&\left[(\qb^\mu\cosh w{-}i\epsilon^{\mu\nu\alpha\beta} \qb_\nu u_{1\alpha} u_{2\beta}) e^{-i\qb\cdot( b+i {a})}\right] \,.
\end{align}
As one can see we have recovered eq.(\ref{eq:FTimpulse}): importantly, we identify the shift in Kerr solution explicitly with the exponentiation of $\frac{s}{m}$ for spinning particles in the large spin limit! Evidently the shift $b \rightarrow b + ia$ arises because of the exponential structure of minimally coupled amplitudes, and the Fourier factor $e^{i \qb \cdot b}$ in expressions for observables in terms of amplitudes.

%%%%%%%%%%%%%%%%%%%%%%%%%%%%%%%%%%%%%%%%%%%
\subsection{Impulse for Kerr black hole }
%%%%%%%%%%%%%%%%%%%%%%%%%%%%%%%%%%%%%%%%%%%
The leading order impulse for a spinning black hole was derived to all orders in spin by Vines~\cite{Vines:2017hyw}. In impact parameter space it takes the form:
\begin{align}\label{GraviImpulse}
\Delta p_1^\mu &=-2G m_1 m_2 \Re \left[(\cosh 2w\, \eta_{\mu\nu}{+}2 i \cosh w\, \epsilon_{\mu\nu\rho\sigma}u_1^\rho u_2^\sigma)\right.\nonumber\\
&\quad\quad\left.\frac{(b+i\Pi a)^\nu}{\sinh w(b+i\Pi a)^2}\right]
\end{align}
This result follows straightforwardly from our $\sqrt{\rm Kerr}$ discussion, by simply ``squaring" the $x$-factors in eq.(\ref{Factor}), and replacing ${\sqrt2}e\rightarrow {- \kappa/2}$. The result is just a factor of two for the rapidity factor in eq.(\ref{Rapid})
\begin{align}
\Delta p_1^\mu   & ={-}i 2\pi G m_1 m_2\int \! \hat{d}^4\qb\; \hat{\delta}(\qb\cdot u_1)\hat{\delta}(\qb\cdot u_2)e^{{-}i\qb\cdot b} \frac{\qb^\mu }{\qb^2} \nonumber\\
&\left((\cosh 2w{+}\sinh 2w) e^{\qb\cdot{a}}{+}(\cosh 2w{-}\sinh 2w) e^{{-}\qb\cdot{a}}\right)
\end{align}
Again, using the identities for $\sinh w$ we derived previously, we can rewrite
\eq
 \qb_\mu \sinh 2w=2  \qb_\mu \cosh w \sinh w = i2\cosh w\,\epsilon_{\mu\nu\alpha\beta} \qb^\nu u_1^\alpha u_2^\beta
\eqe
and thus
\begin{align}
\Delta p_1^\mu &= {-}4\pi G m_1 m_2\Re\int \! \hat{d}^4\qb\; \hat{\delta}(\qb\cdot u_1)\hat{\delta}(\qb\cdot u_2)\frac{ie^{{-}i\qb\cdot( b+i {a})}}{\qb^2}  \nonumber\\
&\quad\quad (\qb^\mu\cosh 2w{+}i2\cosh w\,\epsilon_{\mu\nu\alpha\beta} \qb^\nu u_1^\alpha u_2^\beta) \,.
\end{align}
To compute the Fourier integrals, note that $q \cdot b = q \cdot \Pi b$ on the support of the integral, where $\Pi$ is the projector onto the space orthogonal to $u_1$ and $u_2$. With this replacement, the Fourier integral is
straightforward to compute\footnote{For example, see page 33 of~\cite{Kosower:2018adc}},  and the result is
\[
\Delta p_1^\mu = {-}\frac{2m_1 m_2 G}{\sinh w}\Re  \left[  \frac{\cosh 2w\,b_\perp^\mu{+}i2\cosh w \epsilon^{\mu\nu\alpha\beta} u_{1\alpha} u_{2\beta}b_{\perp\nu}}{b_\perp^2} \right] \,,
\label{eq:finalImpulse}
\]
where $b_\perp=\Pi(b+i a)$, in agreement with eq.(\ref{GraviImpulse}).
%%%%%%%%
\vspace{-0.3cm}

\vspace{-0.4cm}

%%%%%%%%%%%%%%%%%%%%%%%%%%%%%%%%
\section{Discussions and Conclusions}
%%%%%%%%%%%%%%%%%%%%%%%%%%%%%%%%%%%%%%%%
In this paper we demonstrated that the exponentiation induced in taking the large-spin limit of minimally coupled spinning particles, precisely maps to the Newman-Janis complex shift relating the Schwarzschild and Kerr solutions in position space.

Note that these are very general features, applying to a wide range of observables (eg the total change in spin of a particle during a scattering event~\cite{angularimpulse}) and in a wide range of theories, including Einstein gravity. Moreover, while we have described the situation in detail at lowest order for the impulse, one can compute the impulse to all orders using scattering amplitudes. The only three point vertex available for particles moving in the static background field is the \rootKerr amplitude of equation~\eqref{eq:RootKerrAmplitude}. Thus the replacement must hold to all orders, as well as for its gravitational counterpart.

There is still much to learn by studying classical gravity from the point of view of on-shell methods. We need to learn more about amplitudes for particles with large spins in order to understand the dynamics of Kerr black hole scattering (not just probe scattering) in more detail. Moreover, the interplay between the double copy, massive particles, and Einstein gravity needs to be explored in more detail~\cite{henrik}, especially in light of recent difficulties~\cite{Plefka:2019hmz}.

It will be interesting to explore the correspondence to other solutions where either complex shifting or double copy relations hold. This includes the shifting relation between Kerr-Newman and Reissner-Nordstrom, as well as the double copy relation between dyons and the Taub-NUT solution~\cite{Luna:2015paa}. We leave this for future work.

Finally, in this note we have focused on understanding the three-point couplings of Kerr black holes in the simplest and most physical way, involving the scattering of probe particles in asymptotically Minkowski spacetime. We did not directly consider the on-shell three-particle scattering. Indeed, it is a standard (and important) fact of basic kinematics that the 3-particle amplitude is never ``on-shell'' in asymptotically Minkowski space. Instead the three-particle amplitude makes sense for general complex momenta, and also for real momenta, not in $(3,1)$ but in $(2,2)$ signature. Clearly, the complexification associated with the Kerr solution is begging for a formulation in $(2,2)$ signature, where an even more direct computation of the three-particle amplitude should be possible. Given the important role of $(2,2)$ signature physics in many other aspects of four-dimensional scattering amplitudes, this may well be more generally a fruitful avenue of exploration for future work.

%%%%%%%%%%%%%%%%%%%%%%%%%%%%%%%%%%%%%%%%%%%%%%%%%%%%%
\section{Acknowledgements}
%%%%%%%%%%%%%%%%%%%%%%%%%%%%%%%%%%%%%%%%%%%%%%%%%%%%
We thank Andr\'es Luna, Lionel Mason, Ricardo Monteiro and Justin Vines for useful discussions. NAH is supported by DOE grant de-sc0009988, YTH is supported by MoST Grant No. 106-2628-M-002-012-MY3, while DOC is supported by the STFC grant ``Particle Theory at the Higgs Centre''. YTH and DOC would like to thank Simons Foundation for its support for the ``Amplitudes meets cosmology" workshop, during which this work was done.

%{99}
\end{document}